 \documentclass{emulateapj}
 \usepackage{colordvi}
\newcommand{\dndz}{$d\cal{N}$/$d z$}
\newcommand{\dndn}{$d\cal{N}$/$dNdz$}
\input epsf

\begin{document}
\title{HOT BARYONS AND THE DISTRIBUTION OF METALS IN THE INTERGALACTIC MEDIUM}
\author{
JASON TUMLINSON\altaffilmark{1} \&
TAOTAO FANG\altaffilmark{2}}

\altaffiltext{1}{Department of Astronomy and Astrophysics,
                 University of Chicago,
                 5640 S. Ellis Ave., Chicago, IL 60637}
\altaffiltext{2}{Chandra Fellow, Astronomy Department, 601 Campbell Hall,
University of California, Berkeley, CA 94720}

\begin{abstract}
We use the observed number and column-density distributions of
intergalactic \ion{O}{6} absorbers to constrain the distribution of
metals in the low-redshift intergalactic medium (IGM). In this simple
model the metals in the \ion{O}{6} absorbers are assumed to be produced
in and propagated from low-redshift galaxies drawn from a real sample,
in this case the Sloan Digital Sky Survey (SDSS). This model can explain
the observed \dndz\ of metals seen in \ion{O}{6} absorbers if these
metals are dispersed out to $\sim 200$ kpc by galaxies down to $L \sim
0.01 - 0.1 L_r^*$. Massive galaxies ($L \sim L_r^*$) by themselves cannot
provide the necessary enrichment unless they can enrich volumes out to
$\gtrsim 0.5 - 1$ Mpc. Alternatively, our model allows an estimate of
the fraction of \ion{O}{6} absorbers directly caused by galaxies rather
than hot IGM. With this assumption we explore the possible connections
between the intergalactic \ion{O}{6} absorbers and the known populations
of highly-ionized high-velocity clouds (HVCs) surrounding the Milky Way.
Our model predicts that more sensitive, complete surveys optimized to
uncover weaker \ion{O}{6} absorbers will find the tentative turnover below
$\log N$(\ion{O}{6}) $\sim 13.5$ to be a real effect resulting from the
apparently limited volumes over which galaxies can enrich the IGM. If
so, it would indicate that metals are not as widespread throughout the
low-density IGM as they are assumed to be in cosmological simulations
of the WHIM.  \end{abstract}


\section{INTRODUCTION}

Substantial populations of intergalactic QSO absorption-line systems
have been detected in highly-ionized oxygen (\ion{O}{6}) with the Hubble
Space Telescope (HST) and the {\em Far Ultraviolet Spectroscopic Explorer}
({\em FUSE}). Because they are probably hot ($T \gtrsim 10^5$~K), these
absorbers may represent a portion of the ``missing baryons'' that are
predicted (Cen \& Ostriker 1999; Dav\'{e} et al. 1999) to arise at low
redshift from continuous structure formation in the warm-hot intergalactic
medium (WHIM). Early surveys of these absorbers with HST (Tripp, Savage,
\& Jenkins 2000; Savage et al.~2002, Tripp 2005) found unexpectedly
high number densities per unit redshift, \dndz\ $\sim 16$, down to 30 -
50 m\AA\ sensitivity. The recent {\it FUSE} study by Danforth \& Shull
(2005, hereafter DS05) has refined these estimates of \dndz\ and measured
their column density distribution, \dndn. DS05 assumed uniform 0.1 solar
metallicity ([O/H] $= -1.0$) and $T = 10^{5.5}$ K to estimate the total
contribution of the \ion{O}{6} absorbers to the baryon density, $\Omega
_{OVI} h_{70}^{-1} = 0.0022 \pm 0.0003$, where $\Omega _b h_{70}^{2}
= 0.024 \pm 0.001$ (Spergel et al. 2003)\footnote{We assume a Hubble
constant $H_0 = (70$ km s$^{-1}$ Mpc$^{-1}$) $h_{70}$.}.

The {\it FUSE} data cannot measure the gas metallicity precisely but
they allow for estimates of typical values. DS05 detect 40 \ion{O}{6}
absorbers which span 2.5 dex in $N$(\ion{H}{1}) but only 1.5 dex in
$N$(\ion{O}{6}). By combining photoionization and collisional ionization
equilibrium (CIE) models with a theoretical relation between IGM
density and $N_{HI}$, DS05 derive a typical metallicity of $Z_O = 0.09
Z_{\odot}$. We derive a conservative lower limit [O/H] $= -2.3 \pm 0.05$
from a pure CIE model that matches the broad locus of points in Figure 1
of DS05, assuming the maximum fractional ionization, $f_{\rm O VI} = 0.2$,
of \ion{O}{6} at $T = 10^{5.45}$ K. However, according to Equation 6 of
DS05 this metallicity limit implies $\Omega _{OVI} h_{70}^{-1} = 0.044$,
close to the total cosmic baryon density. Because a major portion of the
WHIM is thought to lie in hotter gas accessible only to X-ray telescopes
(Fang et al. 2002; Mathur, Weinberg, \& Chen 2003; McKernan et al. 2003;
Nicastro et al. 2005), it is likely that the true metallicity of these
absorbers lies closer to the fiducial value [O/H] $= -1.0$ assumed by
DS05. Metallicity [O/H] $\gtrsim -1.0$ in $\sim 30$\%  of the detected
Ly$\alpha$ clouds (DS05) is too high and too common to have been in
place by $z \sim 3$, where Schaye et al. (2003) find filling factors
of 2 -- 10\% for gas with [C/H] $\geq -2.0$. The \ion{O}{6} absorbers
therefore imply significant later enrichment by galaxies. Conversely,
if the metallicities of these absorbers lie near our lower limits,
a portion of the sample would need to be removed to limit their cosmic
density. Thus the mean metallicity, the dispersion in the metallicity, and
the possibility of other mechanisms for producing \ion{O}{6} are essential
to understanding the contribution of \ion{O}{6} absorbers to $\Omega _b$.

To produce detectable \ion{O}{6} absorption, an intervening cloud must
(1) have the proper ionization conditions to ionize oxygen to \ion{O}{6},
and (2) have sufficient metallicity to make the \ion{O}{6} lines strong
enough to detect. The first condition can be met in CIE gas produced in
shocks (Shull \& McKee 1979) or in hot-cold interfaces (Fox et al. 2004;
Indebetouw \& Shull 2004, and references therein), though for some IGM
conditions photoionization by a metagalactic ionizing background (Haardt
\& Madau 1996) can also produce \ion{O}{6}. The ionization mechanism
need not be caused by, nor occur near, a galaxy, but in the standard
picture of structure formation metals must be created in galaxies. When
absorber/galaxy relationships are considered, there is no necessary
connection between galaxies and the {\it ionization} of the gas, while
there is a necessary connection between galaxies and the {\it metallicity}
of the gas. Thus even if the \ion{O}{6} absorbers fill only a portion
of the IGM hot-gas budget, they can still constrain the distribution of
metals. In this paper we study the dispersal of IGM metals by galaxies
using only {\it observables} and simple geometric arguments. Our
model does not attempt to explain the shock-heating or photoionization
mechanisms that cause or maintain the ionization of \ion{O}{6}, but
rather the metals that must have some link to nearby galaxies.

Our model has significant advantages over simulations (Cen et
al.~2002; Fang \& Bryan 2001; Chen et al.~2003; Yoshida et al. 2005)
or semi-analytic calculations (Furlanetto et al.~2005) that predict
\dndz\ from {\it a priori} models. These models either do not treat
metal distribution in a self-consistent way, by adopting a uniform
or density-dependent metallicity, or they use complicated models
to calculate ionization structures and obtain reasonable \ion{O}{6}
ionization fractions. By contrast, our model avoids all these complicated
issues. By decoupling these entangled physical processes, we can then
focus on key issues, namely the volume that can be enriched by galaxies,
and the minimum luminosity of the galaxies that can enrich the nearby
IGM. Our method also allows us to evaluate the hypothesis that the
intergalactic \ion{O}{6} absorbers are related to the \ion{O}{6} HVCs
seen in the Galaxy. Combining some simple geometrical arguments we
derive the expected column density distribution, \dndn. When compared
with observations, our simple model provides significant constraints on
the distribution of metals and their connection with galaxies.

\section{QSO/Galaxy Pairs and IGM Metal Enrichment}

Our model compares the distribution of IGM metals expressed by \dndz\ to
the incidence of QSO-galaxy ``pairs'' expected per redshift interval. We
calculate the total number of absorbers $dn$ that can be encountered
by one random sightline along a distance $dl$ to a background QSO. The
spatial density of halos with absolute magnitude between $M_r$ and
$M_r+dM_r$ is $(dn_h/dM_r)dM_r$, where $dn_h/dM_r$ will ultimately
be determined by SDSS.  Assuming each halo can enrich the surrounding
volume to a radius of $R$, the total number of such intercepted absorbers
within a distance of $dl$ along the line of sight (LOS) toward background
quasars then is
\begin{equation}
\frac{dn}{dl} = f \pi R^2 \int_{-\infty}^{M_r^{faint}} \frac{dn}{dM_r}dM_r.
\end{equation}
Here $f$ is the LOS covering fraction, representing the chances of
intercepting absorbers within each halo. In the next section we will see
that $f \sim 1$ when we relate the local \ion{O}{6} high velocity clouds
(HVCs) and the intergalactic \ion{O}{6} absorbers.

In principle the integral in the last term of Equation 1 could be
stated in terms of mass and derived from, e.g. a Press-Schechter
function, but in practice we use observations. We begin with the
SDSS Data Release 3 catalog of QSOs (all spectroscopically confirmed)
and galaxies\footnote{Available at http://skyserver.pha.jhu.edu.}. To
maximize the size of the sample and therefore provide good statistics,
we use all SDSS photometric sources determined to be galaxies by
photometric selection, then calculate photometric redshifts, $z_{\rm
photo}$, from the SDSS $ugriz$ photometry using the software of Blanton
et al. (2003). These redshifts have typical errors $\sigma _z = 0.02 -
0.04$, which make individual pairs quite uncertain but which wash out
in the average. For each QSO sightline, we assign to all galaxies within
42$'$ an impact parameter $b$ to the QSO sightline at the galaxy $z_{\rm
photo}$. We require $z_{\rm photo} = 0.02 - 0.2$ to minimize spurious
detections on the low end and incompleteness at the high end. We correct
for incompleteness in faint galaxies at the higher redshifts by assigning
each galaxy (or magnitude bin of galaxies) a unique pathlength $\Delta
z$ based on its absolute magnitude and the SDSS flux limits. Then,
for a given spherical ``halo'' or ``enrichment'' radius $R$ and a
LOS covering fraction $f$, we can derive the total number and \dndz\
of pairs, with cuts for any galaxy property (i.e., color, morphology,
luminosity) measurable from the $ugriz$ photometry and $z_{\rm photo}$.

\begin{figure}[t]
\centerline{\epsfxsize=1.1\hsize{\epsfbox{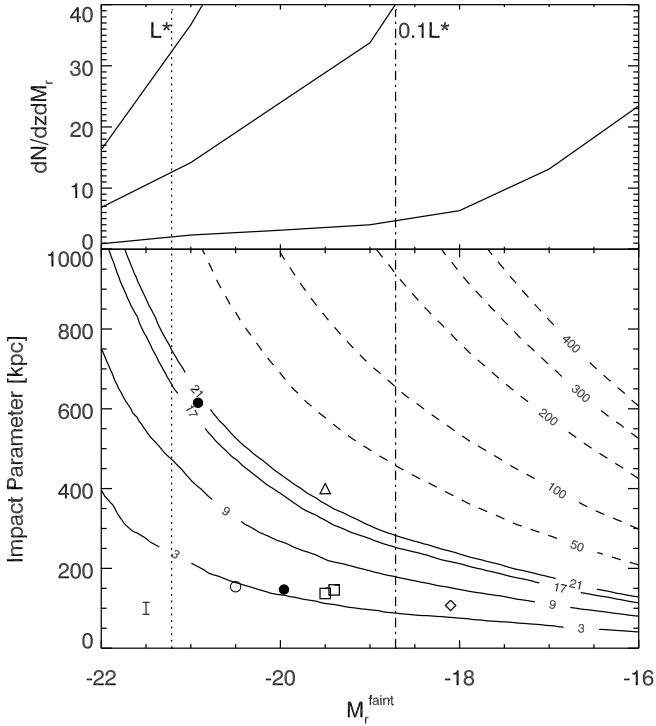}}}
\figcaption{Upper panel: Differential distribution of \dndz\ as a function
of galaxy luminosity, in 1 mag bins. From top, the lines correspond to
$R =$ 500, 300, and 100 kpc. Lower panel: As a function of $b$ and SDSS
$r$-band absolute magnitude, contours of constant cumulative \dndz\
obtained in the fiducial model ($f = 1$). Clearly dwarf galaxies must
enrich their environments for the observed number density of metal-line
absorbers to be obtained. The vertical lines mark $L^*_r$ and 0.1$L^*_r$
from the Blanton et al. (2003) SDSS luminosity functions. Solid contours
for \dndz\ = 3, 9, 17, and 21 correspond to equivalent width limits
$W_{\lambda}$ = 100, 50, 30, and 10 m\AA, respectively (DS05). The
dashed contours show model predictions for higher \dndz. The error
bar at the lower left shows the approximate vertical range associated
with the errors bars in the observed \dndz. The symbols mark the
luminosities and impact parameters of nearest-neighbor galaxies for
seven well-studied QSO absorbers. These are: open circle, H1821+643
(Tripp et al. 2001); filled circles, PG0953+415 (Savage et al. 2002);
open triangle, PKS2155-304 (Shull, Tumlinson, \& Giroux 2003); diamond,
PKS0405-419 (Prochaska et al. 2005); open squares, PG1211+143 (Tumlinson
et al. 2005). \label{dndzfig}} \vspace{-0.2in} \end{figure}

Figure~\ref{dndzfig} shows differential and cumulative distributions
of \dndz\ achieved by galaxies of varying $L_r$, given a uniform impact
parameter $b$ and $f = 1$. The key measurements are \dndz\ = 3, 9, 17, 19,
and 21 for $W_{\lambda} \geq$ 100, 50, 30, 15, and 10 m\AA, respectively
(DS05). This mapping between galaxy populations and absorber incidence
is the major result of this study. Spherical enrichment zones must be
quite large, $b \gtrsim 0.5$ Mpc, and uniform ($f = 1$) for enrichment
by massive galaxies ($L \gtrsim L_r^*$) to explain the observed numbers
of \ion{O}{6} absorbers. For $b \sim 200$ kpc, low-mass galaxies ($L
\sim 0.1 L_r^*$) can reproduce the observed \dndz. To obtain \dndz\
= 21 with dwarf galaxies only, we require uniform $R = 145$ kpc for
$M_r = -19$ to $-16$ and $R = 160$ kpc for $M_r = -18$ to $-16$. In
passing we note our model can also constrain $R$ for \ion{H}{1} halos
corresponding to high-$N$(\ion{H}{1}) Ly$\alpha$ clouds, i.e. $R = 50$
kpc at $L \gtrsim L_r^*$ is needed to reproduce \dndz\ = 1 for Lyman
limit systems (Keeney et al. 2005).

DS05 find a power-law dependence of \dndz\ on $N_{OVI}$, $d\cal{N}$/$
d N_{\rm OVI} \propto N_{\rm OVI}^{-2.2}$ for absorbers with $\log
N$(\ion{O}{6}) $\geq 13.4$. The behavior of \dndn\ at low $N$ is of
critical importance to estimating $\Omega _{\rm OVI}$. However, below
$\log N$(\ion{O}{6}) $= 13.4$, the observed numbers of \ion{O}{6} absorbers
drop below the power-law fit , and well below the numbers expected from
simulations that assume uniform or overdensity-dependent metallicity in
the WHIM (Figure~\ref{dndnfig}). DS05 cannot say whether the apparent
turnover below $\log N$(\ion{O}{6}) $\simeq$ 13.5 is real or due to survey
incompleteness below $\sim 50$ m\AA. Integrating this power law down to
$\log N$(\ion{O}{6}) = 13.0 (12.5 m\AA) yields \dndz\ = 106. This high
number density requires [O/H] $\gtrsim -2$ over $\sim 600$ kpc spheres
for $L \gtrsim 0.1 L_r^*$ (Figure 1), an unlikely outcome. We believe
it is more likely that the observed turnover below $\log N$(\ion{O}{6})
$\simeq$ 13.5 is real and not a consequence of incompleteness in the
DS05 survey. This prediction could be verified by a survey with more
pathlength coverage down to $\log N$(\ion{O}{6}) = 13.0 (12.5 m\AA). If
this deficit relative to theoretical expectations is real, our model
suggests that metals are not as widely or uniformly distributed throughout
the low-density IGM as simulations assume or find them to be.

QSO absorption-line studies of the dispersal of IGM metals generally offer
conflicting guidance because they suffer from the practical difficulty
of identifying the faintest galaxy with which a given absorber might
be associated. These studies cannot generally exclude the presence of a
faint galaxy nearer the sightline than a known bright one, and so they
tend to overestimate the dispersal of metals. This effect is seen in the
data points in Figure 1, which show $M_r$ and $b$ of nearest-neighbor
galaxies to \ion{O}{6} absorbers with galaxy surveys of varying depth and
sky coverage. There is a general trend of metals spread to $\sim 100
- 200$ kpc, with some notable exceptions. The $z = 0.06807$ absorber
toward PG0953+415 (Savage et al. 2002)  suggests $\gtrsim 0.2$ solar
enrichment to $\sim 600$ kpc, but this field has not been surveyed much
below $L^*_r$. The PKS2155-304 \ion{O}{6} absorbers (Shull, Tumlinson,
\& Giroux 2003) provide the best case for metal dispersal to $\simeq
400$ kpc, because this field has been surveyed to $L \sim 0.01L^*_r$,
but the absorbers appear to be associated with a galaxy group that
may complicate constraints on metal dispersal. Chen et al. (2001)
systematically studied \ion{C}{4} near 50 galaxies over a range of
luminosity and found $f = 0.67$ at $b \leq 100$ kpc and $f = 0.06$ at $
b \geq 100$ kpc, with typical limiting $W_{\lambda} \sim 100$ m\AA. This
result does not rely on a post-facto galaxy survey, and so provides more
robust evidence that luminous galaxies have associated metals out to $b
\sim 100$ kpc. Stocke et al. (2004) found a post-starburst dwarf with
$M_B = -13.9$ associated with a 6\% solar metallicity absorber in the
sightline to 3C~273. This galaxy, while starbursting, apparently drove
a metal-enriched superwind $71h_{70}^{-1}$ kpc into the IGM. Thus there
is evidence that metal enrichment spreads to $\sim 100$ kpc from all
galaxies, but evidence for further propagation is less clear.

We stress that the known \dndz\ of \ion{O}{6} absorbers represents the
minimum budget of metals that must be explained in terms of enrichment
by galaxies. There are surely other reservoirs of metals that meet
condition 2 above (detectable metallicity) but not 1 (conditions that
produce \ion{O}{6}). Any difficulty associated with distributing the
oxygen seen in \ion{O}{6} can only be compounded by the addition of
these other reservoirs (e.g., photoionized gas in \ion{C}{4} systems)
to the budget of IGM metals. Finally, Figure 1 shows results for $f =1$
and so presents the most optimistic case of enrichment.

\section{Intergalactic O VI and High-Velocity Clouds}

Our model can also interpret the intergalactic O VI absorbers as analogs
of the local highly-ionized high-velocity clouds (Sembach et al.~2003,
hereafter S03). To do so, we must first derive a mapping between the
covering fraction, $C$, of HVCs with column density $N$ as seen along
radial sightlines out from the Galaxy, to $f$, the covering fraction along
a LOS that follows a chord through an intervening galaxy halo. The key
assumption here is that the HVCs are not {\it unique}, i.e., an observer
located in another galaxy similar to our Galaxy should be able to detect
similar HVCs with a similar pattern of column density distribution. Let
us first consider HVCs with column densities between $N_j$ and $N_j +
\Delta N_j$, and with a covering fraction $C_j$ -- the total covering
fraction $C$ is the sum of $C_j$. Assuming there are a total of $T_j$
such absorbers uniformly distributed within the halo (of a galaxy or
a group), and each absorber has a spherical geometry with radius $d$,
we find $T_j$ and the spatial number density of the absorbers within
the halo $n_{j}$ are
\begin{equation}
T_j = \frac{4R^2}{3d^2}C_j \; ; \; n_{j} \equiv \frac{T_j}{V} =
\frac{C_j}{\pi d^2 R},
\end{equation} respectively, where  $R$ is the radius of the halo. These
absorbers project onto the sky a total area $A_j = T_j \pi d^2$, such
that their covering fraction along the LOS is
\begin{equation}
f_j = \frac{A_j}{\pi R^2} = \frac{4}{3} C_j.
\end{equation}

\begin{figure}[!t]
\centerline{\epsfxsize=1.0\hsize{\epsfbox{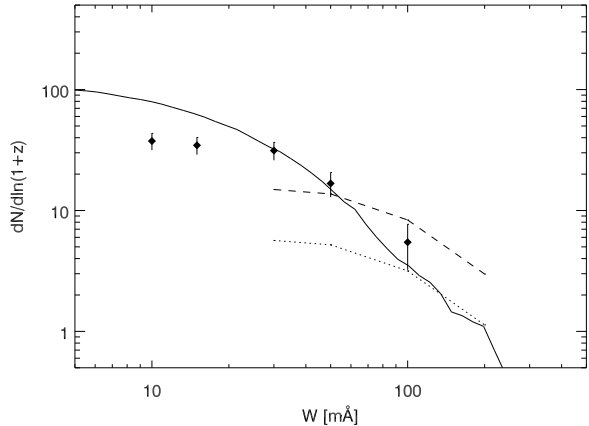}}} \figcaption{Cumulative
\dndz\ of \ion{O}{6} absorbers as a function of $W_\lambda$ for the
DS05 data points (filled diamonds) and theory. The solid line marks
the prediction for the WHIM simulation of Fang \& Bryan (2001), with
an overdensity-dependent metallicity. The dotted and dashed lines mark
the present model for \dndz\ with column-density dependent covering
fractions $f_j = 4C_j / 3$ from S03, for luminosity-impact parameter
pairs $(-19,100)$ and $(-16,80)$, respectively. Note that HVC-type
\ion{O}{6} absorbers may make significant contributions to the observed
numbers, especially if they arise from dwarf galaxies. \label{dndnfig}}
\vspace{-0.2in}\end{figure}

We derive all quantities from observation - $C_j$ from Table 4 of S03
and the spatial density of galaxies from SDSS. The total LOS covering
fraction $f$ then is $f = \Sigma_j f_j = (4/3)C$. Adopting the value of
$C$ from S03, we find $f$ ranges from $\sim 0.64$ to $\sim 1.1$ down
to 30 m\AA,\, which justifies the value of 1 we adopt in the previous
section. Our results appear in Figure~\ref{dndnfig}, where we show that
HVC-like \ion{O}{6} absorbers can make a significant contribution to
the observed numbers if massive galaxies have large halos, or if all
galaxies down to $M_r \sim -16$ ($\sim 0.01 L_r^*$) have \ion{O}{6} HVCs
out to $\sim 100$ kpc. Highly ionized gas can arise in galaxy superwinds
(Heckman et al. 2001), in gas tidally stripped from galaxies (Cox et
al. 2004), and in accreting gas (Maller \& Bullock 2004; Birnboim \&
Dekel 2002). The $M_B = -13.9$ dwarf 71$h_{70}^{-1}$ kpc away from
a metal-enriched absorber toward 3C~273 (Stocke et al. 2004) is one
example at the faint end of the luminosity function. The {\em FUSE}
studies of Galactic \ion{O}{6} HVCs (Sembach et al. 2003; Fox et al. 2004;
Collins, Shull, Giroux 2004) show that \ion{O}{6} can arise in the extended
halo regions of larger galaxies ($\sim 100$ kpc) and not necessarily
correspond to the WHIM.   From Figure~\ref{dndnfig}, it appears that if
these processes have high $f$ out to $\sim 100$ kpc around galaxies down
to $\sim 0.01 L^*_r$, they may contribute \dndz\ $\sim 5$ to the overall
population of \ion{O}{6} absorbers. Given the potentially important
contributions of \ion{O}{6} absorbers to $\Omega_{\rm WHIM}$, their utility as
pointers to the hotter WHIM seen in X-ray absorption, and considering
the many mechanisms that can produce \ion{O}{6} near galaxies, it is
important to obtain an unbiased observation of the incidence of \ion{O}{6}
and metals within $100 - 200$ kpc of galaxies over a range of mass. Until
such an observation is made, we cannot know how to correct the inferred
baryon densities for these contaminating effects. Our observations can
also guide the design of such studies by constraining the fields over
which galaxy surveys must be executed.

\section{CONCLUSIONS}

Based on our model of the distribution of IGM metals and the observed
numbers of IGM \ion{O}{6} absorbers, we draw the following conclusions:

\begin{enumerate}
\item Unless they eject metals to $\sim 0.5 - 1$ Mpc, luminous galaxies
cannot explain \dndz\ for metal-line absorbers in the low-$z$ IGM. If
low-mass galaxies ($\gtrsim 0.01 - 0.1 L^*_r$) disperse metals to $\sim
200$ kpc, \dndz\ for the \ion{O}{6} absorbers can be explained.\\
\item Uniform enrichment to $\sim$ 1 Mpc is implied by integrating
the power-law dependence of \dndn\ (DS05) to $\log N$(\ion{O}{6}) =
13.0. Thus the turnover seen by DS05 in \dndn\ below $\log N$(\ion{O}{6})
$\simeq 13.5$ (50 m\AA) is more likely a real effect than a consequence
of survey incompleteness. A survey sensitive to $W_{\lambda} = 12.5$
m\AA\ could confirm this result and further constrain the dispersal of
IGM metals. Conversely, additional absorbers below $\log N$(\ion{O}{6})
$\simeq 13.5$ would imply that metals are quite widespread.
\item A high covering fraction of \ion{O}{6} from ``HVCs'' within $100 - 200$
kpc from galaxies can contribute \dndz\ = $5 - 10$, a non-negligible
portion of the observed total that would not contribute to $\Omega
_{\rm WHIM}$.

\end{enumerate}

\acknowledgements

Thanks to Charles Danforth and Mike Shull for sharing their results before
publication and to Charles Danforth and John Stocke for constructive
comments.  We began this work in the hospitality of the Kavli Institute
for Theoretical Physics at the University of California, Santa Barbara,
supported in part by NSF under Grant No. PHY99-07949. J. T. is supported
by the Astronomy and Astrophysics Department at the University of Chicago
and HST GO-9874.01-A. T. F. was supported by NASA through {\sl Chandra}
Postdoctoral Fellowship PF3-40030 issued by the {\sl Chandra} X-ray
Observatory Center, which is operated by the Smithsonian Astrophysical
Observatory for and on behalf of NASA under contract NAS8-39073.

\end{document}